\documentclass{aa}
\usepackage{graphicx}
\usepackage{amsmath,epsfig,rotating,amssymb}

\begin{document}

\def \rot{{\rm {\bf rot} }}
\def \grad{{\rm {\bf grad} }}
\def \div{{\rm div}}
\def \cha{\widehat}
\def \pr{{\it permanent}  regime }
%\textwidth 6.in
%\oddsidemargin 0.in
%\evensidemargin 0.in
%\topmargin 0.in
%\headheight 0.5in
%\headsep 0.2in
%\textheight 9.in
%\parindent 0.2in
%\pagestyle{myheadings}

%\input{psfig.tex}

\author{Hennebelle P. \inst{1}, Audit E. \inst{2} and Miville-Desch\^enes M.-A. \inst{3}}

\institute{Laboratoire de radioastronomie millim{\'e}trique, UMR 8112 du
CNRS, 
\newline {\'E}cole normale sup{\'e}rieure et Observatoire de Paris,
 24 rue Lhomond, 75231 Paris cedex 05, France 
\and  Service d'Astrophysique, CEA/DSM/DAPNIA/SAp, C. E. Saclay,
  F-91191 Gif-sur-Yvette Cedex 
\and
Institut d'astrophysique spatiale, Universit\'e Paris-Sud, Bat. 121, 91405, Orsay, France.}

\offprints{E. Audit, P. Hennebelle  \\
{\it e-mail:} edouard.audit@cea.fr, patrick.hennebelle@ens.fr}   

\title{On the structure of the turbulent interstellar atomic hydrogen. II- First 
comparison between observation and theory}

\subtitle{Are the characteristics of molecular clouds determined early 
in the turbulent 2-phase atomic gas?}

\titlerunning{Comparing simulations and observations of HI}

\abstract
{}
{It is necessary to understand the dynamics of the atomic gas to use complex modeling and 
to carry out detailed comparisons between theoretical models and observations.}
{In a companion paper, we  present high resolution bidimensional  numerical simulations of the 
interstellar atomic hydrogen. Here, we further characterize  these simulations and 
we compare our results with various observations.}
{We give statistics of the column density and velocity along the line of sight and show that 
they compare favorably with observations of high-latitude lines of sight. 
We compute synthetic HI spectra 
and  qualitatively discuss the information that could be inferred if these spectra were observed. 
Finally, we extract 
CNM clouds and study their physical properties finding strong similarities with real  clouds.
In particular, we find that  the clouds follow Larson-type relations, 
i.e $M \propto L^\gamma$, where $\gamma \simeq 1.7$ (we speculate that in 3D, $\gamma \simeq 2.5$)
and $\sqrt{<\delta v^2>} \propto L^{0.4}$.
We also find that the distribution, ${\cal N}(N)$,  of the column density, $N$, of the CNM
structures  formed in the simulation follows ${\cal N}(N) \propto N^{-1.2}$ which is 
marginally compatible with the observational result  obtained by Heiles \& Troland (2005).
From the mass-size relation and the mass spectrum, we derive an exponent 
for the column density distribution close to the value obtained in the numerical 
simulation. }
{We conclude that the simulations reproduce  various observational features reasonably well. 
An important implication suggested by our results is that the  "turbulence" within 
the cold interstellar atomic gas is mainly the result of individual long living cloudlet 
(confined by an external warm medium) motions rather 
than supersonic turbulence within nearly isothermal clouds.
Another important aspect is that the CNM structures produced in the simulation present 
various physical characteristics that are similar to the characteristics of 
the molecular clouds. This 
raises the question as to whether the physical properties  of the   molecular clouds  are 
determined at a very early stage, before  the gas  becomes molecular. }

\maketitle

\section{Introduction}

Since its first use in the 1950s (Ewen \& Purcell, 1951)
the 21 cm transition of atomic hydrogen (HI) has been extensively used
(e.g. Crovisier 1981, Kulkarni \& Heiles 1987, Dickey \& Lockman 1990, 
Joncas et al. 1992)
to study the neutral atomic interstellar gas by numerous authors. Recent data have considerably improved
our knowledge of this medium.
 Heiles \& Troland (2003, 2005), using the Arecibo telescope, have done an extensive 
survey of HI clouds  which has given reliable statistical results.
To directly  probe smaller scales of HI emission, various studies have been done 
using interferometers.  
 Miville-Desch\^enes et al. (2003) and Taylor et al. (2003) use the Dominion Radio Astrophysical 
Observatory (DRAO) interferometer, Stil et al. (2006) use the Very Large Array (VLA)
whereas McClure-Griffiths et al. (2005) use the Australian Telescope Compact Array (ATCA).

The observations of HI have revealed its complex multi-phase nature.  
In HI, cold and dense (CNM) structures are embedded in a warm (WNM) phase,  
 with which there are approximately in  pressure equilibrium,  
as predicted by detailed computations of thermal balance (Field et al. 1969, Wolfire et al. 1995, 2003).
The  observations have also revealed that this multi-phase medium is strongly  turbulent making 
 their interpretations even more
 challenging and indispensable the use of adapted numerical simulations.

Numerous numerical studies of the large scale ISM (e.g. V\'azquez-Semadeni et al. 1995, 
Rosen \& Bregman 1995,
de Avillez \& Breitschwerdt 2005a) and of  the molecular clouds (e.g. MacLow \& Klessen 2004) 
have been performed over the years and  comparisons between observations and simulations
 have been carried out both  for large scale simulations 
(e.g. Ballesteros-Paredes et al. 2002,  de Avillez \& Breitschwerdt 2005b) 
and  for molecular clouds (Padoan et al. 2003, Ossenkopf \& Mac Low 2002). 

It is, however, only recently that numerical simulations 
devoted  to the description of  the warm and the dense atomic gas with a numerical resolution
sufficient to describe the physical scales of the problem (i.e. the CNM structures and the 
Field length) have been performed (Hennebelle \& P\'erault 1999, 2000, 
Koyama \& Inutsuka 2000, 2002, Audit \& Hennebelle
2005, hereafter paper I,  Heitsch et al. 2005, 2006, V\'azquez-Semadeni et al. 2006). 
For this reason, no tight comparison between observations and theory have 
yet been  carried out for HI at scales below 10 pc. 

In a companion paper (Hennebelle \& Audit 2006, here after paper II), 
we present 2D high resolution numerical 
simulations aiming to describe self-consistently a turbulent atomic hydrogen flow 
from scales of few tens of parsec down to hundredth  of parsec.
The high resolution which is necessary to describe properly HI 
 permits to obtain a good statistical description of the flow and to reach 
 small scales close to the size of  various small scale 
structures that have been observed in HI, like the so-called tiny small atomic structure 
(TSAS) and low column density clouds recently observed by Braun \& Kanekar (2005) and 
Stanimirovi\'c \& Heiles (2005). 
Therefore, in spite of the fact that these simulations are only bidimensional, they constitute 
a good starting point to carry out preliminary comparisons with the available HI data. Note that 
the dynamics of spatial scales which is covered by these simulations is about ten times larger than 
what can be done in 3D yet.

 In this paper, we attempt to  further characterize the results of the numerical  simulations 
presented in paper II and to perform comparisons between these simulations  
 and the HI observations obtained in the Millennium  Arecibo 21 centimeter absorption-line 
survey (Heiles \& Troland 2003, 2005) since various interesting statistical informations 
have been extracted from these data.
 We also make a special effort to provide enough material 
 to allow the observers to make easier the comparisons with their observations.
In  Sect.~2,
 we present some statistics of column density and velocity along the lines of sight
since this information is the most direct observable.  
In  Sect.~3, we  present various examples of synthetic atomic  hydrogen 21cm 
line spectra in emission and absorption, showing in each case the corresponding density and 
velocity fields along  the line of sight. We discuss the influence of the convolution
by a Gaussian beam of various width on the synthetic spectra. 
Finally, we focus in Sect.~4 on the CNM structures formed in the simulations 
and their properties trying 
to make the link with some of the cloud properties available from the Millennium survey. 
Section~5 concludes the paper.

\section{Lines of sight}
In paper II, the structure of the flow which results from the numerical 
simulation has been characterized by the probability distribution function 
of the density and pressure as well as by computing the velocity and density
power-spectra and the energy spectrum. However, although very informative, these 
quantities are not easily obtained from the observations (and if so, would  rely on 
various assumptions). On the other hand, column density and average velocity along the 
line of sight are more straightforwardly inferred from observational signals. Thus, 
 we attempt here to  further characterize 
the flow structure by giving some statistics of the column density and the velocity
along the lines of sight. 
Since our experimental setup is strongly 
anisotropic (the flow enters from the left and right box faces and leaves from 
the top and bottom faces),
 we always show the results obtained along the x-axis and along the y-axis. 

\subsection{Line of sight characteristics}

\begin{figure}
\includegraphics[width=9cm,angle=0]{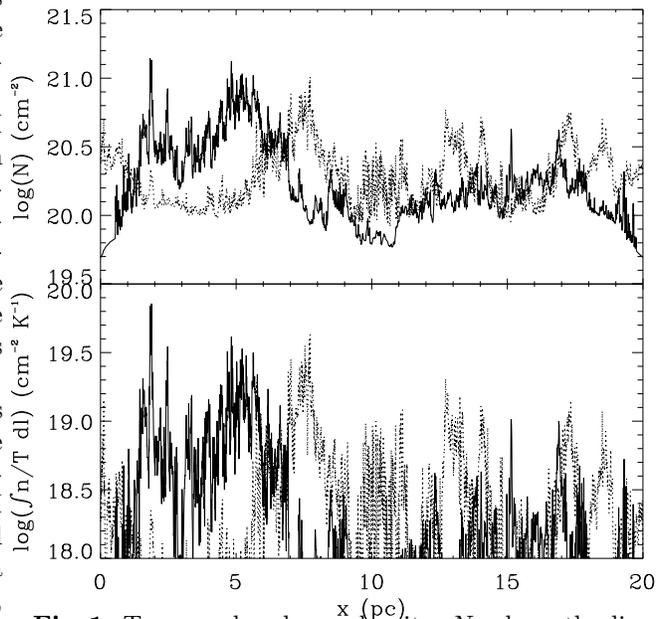}
\caption{Top panel: column density, $N$, along the lines of sight as a 
function of position. Bottom panel: integral of $n/T$ along the line of 
sight.
Full line represents lines of sight parallel to the y-axis, whereas dotted line
represents lines of sight parallel to x-axis.}
\label{col_dens}
\end{figure}

\begin{figure}
\includegraphics[width=9cm,angle=0]{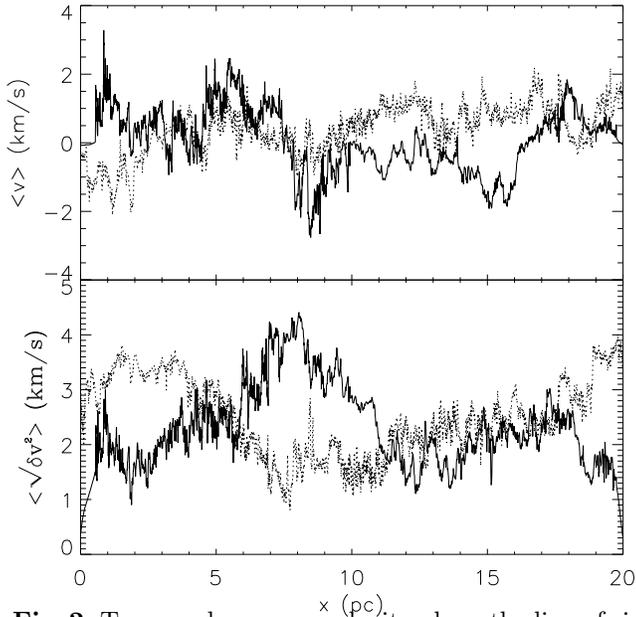}
\caption{Top panel: average velocity along the line of sight.
Bottom panel: velocity dispersion along the line of sight.
Full lines represent lines of sight parallel to y-axis whereas
dotted lines represent lines of sight parallel to x-axis. }
\label{col_vit}
\end{figure}

Top panel of  Fig.~\ref{col_dens} shows the  column density, $N$,
 along the y-axis as a function of the 
x-coordinate (full line) as well as the column density along the x-axis as a function of the 
y-coordinate (dotted line). All the lines of sight available in the simulation, i.e. 10000, are displayed.
 The average value is about $2 \times 10^{20}$ cm$^{-2}$. 
These values are comparable to  several values quoted in the literature for high latitude lines 
of sight (e.g. Heiles \& Troland 2003) and
indicate that the total amount of gas in the simulations is comparable to the total amount of gas along 
these  lines of sight. We note,  however, that only 20-30 $\%$ of the lines of sight quoted by
Heiles \& Troland have such low column density. Our simulation is therefore representative 
of the most diffuse lines of sight. 

Bottom panel of Fig.~\ref{col_dens} shows the integrated quantity $\int n / T dl$ along x and y-axis.
This quantity which traces mainly the CNM is directly obtained by  HI 21cm line  absorption spectra.
An estimate of the CNM column density can be directly obtained by multiplying this quantity by 
the CNM temperature which is about 50 K. Again the values obtained are very similar to the values quoted 
by Heiles \& Troland (2003) for the lines of sight having total HI column densities of about $2 \times 10^{20}$ cm$^{-2}$.

Top panel of Fig.~\ref{col_vit} shows the average velocity along the line of sight defined as:
\begin{eqnarray}
< v_0 > = {\sum \rho v \over \sum \rho}, 
\label{velo_line}
\end{eqnarray}
whereas bottom panel of Fig.~\ref{col_vit} shows the velocity dispersion along the line of sight:
\begin{eqnarray}
<\delta v ^2> = {\sum \rho (v-v_0)^2 \over \sum \rho}.
\label{velo_disp_line}
\end{eqnarray}
The mean velocity is close to zero (0.05 km/s for lines of sight parallel to y-axis and 
0.39 km/s for lines of sight parallel to x-axis) although significant fluctuations (up to 2 km/s) 
are present, the standard deviation is 0.9 and 0.7 km/s for lines of sight parallel to y and x-axis, 
respectively. 
The velocity dispersion along the lines of sight is significantly larger. The mean value is
 about 2.29 and 2.5 km/s along y and 
x-axis respectively, whereas the standard variation is about 0.7 km/s in both cases. Interestingly, 
adopting a CNM sound speed of 0.7 km/s,   
 these values correspond to a Mach number of about $ {\cal M} = 2.5$ km/s / 0.7 km/s $\simeq 3.5$ which is
similar to the values obtained for the CNM clouds by Crovisier (1981) and Heiles \& Troland (2003).

\subsection{Line of sight statistics}

\begin{figure}
\includegraphics[width=9cm,angle=0]{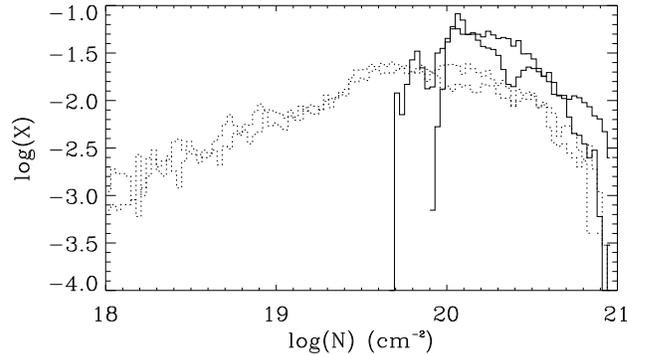}
\caption{Histogram of column density.
The full lines represent the total column density along x and y-axis 
whereas the dotted lines represent the column density (along 
x and y-axis) of CNM only.}
\label{hist_col_dens}
\end{figure}

\begin{figure}
\includegraphics[width=9cm,angle=0]{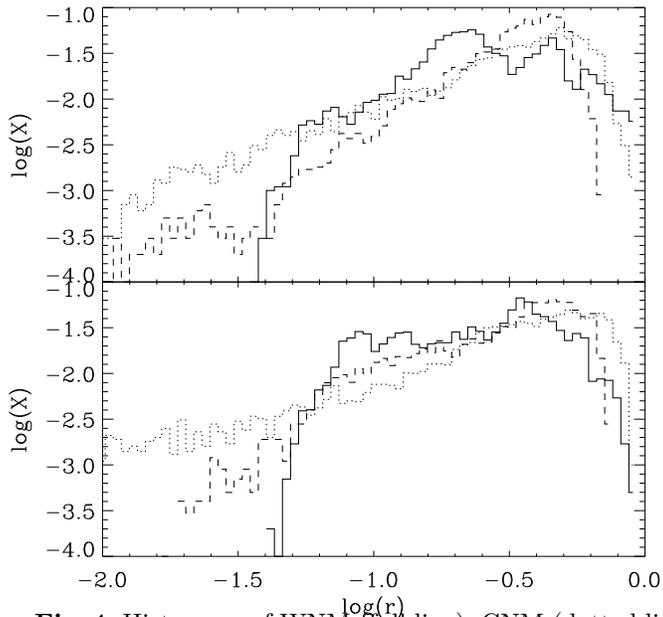}
\caption{Histogram of WNM (full line), CNM (dotted line) and thermally unstable gas (dashed line)
 column density fraction.
Top panel is for lines of sight parallel to  y-axis 
whereas bottom panel is for lines of sight parallel to x-axis. }
\label{rap_col_dens}
\end{figure}

\begin{figure}
\includegraphics[width=9cm,angle=0]{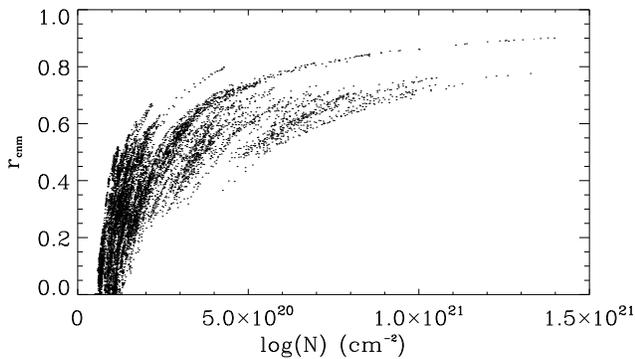}
\caption{Fraction of CNM column density as a function of 
total column density. }
\label{fract_cnm}
\end{figure}

\begin{figure}
\includegraphics[width=9cm,angle=0]{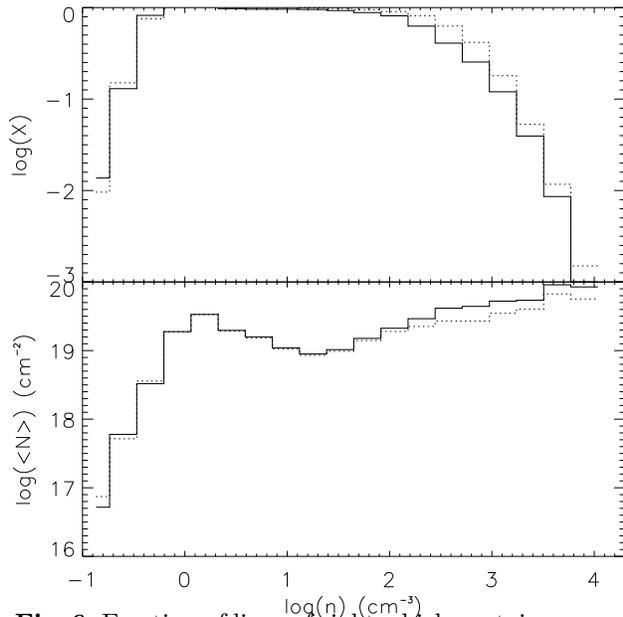}
\caption{Fraction of lines of sight which contain a gas particle of 
a given gas density (upper panel) and mean column density along the line
of sight (bottom panel) of gas having a given density 
(only lines of sight which contain
a fluid particle of the corresponding density are taken into account).
Full lines represent lines of sight parallel to the y-axis whereas dotted lines
represent lines of sight parallel to x-axis.}
\label{line_sight}
\end{figure}

Here, we present statistics of lines of sight. 
Figure~\ref{hist_col_dens} displays the histogram of total column density along x and y-axis (full lines)
and of CNM (defined as gas having temperature lower than 200 K) column density (dotted lines). 
Although our simulation described a coherent region of 20 pc, whereas the lines of sight of the 
Millennium survey are not correlated, it is worth doing some comparison with Fig.~7 of Heiles \& Troland (2003). 
Their first panel shows that CNM column density peaks at $\simeq 1.5 \times 10^{20}$ cm$^{-2}$ close to our 
case ($\simeq 1 \times 10^{20}$ cm$^{-2}$). We find that about 20$\%$ of lines of sight have 
a small ($< 2 \times 10^{19}$ cm$^{-2}$) column density of CNM, whereas Heiles \& Troland  find 
about 30 $\%$  of such lines of sight (first panel of their Fig.~7). 
They find that the total HI column density peaks at about $2 \times 10^{20}$ cm$^{-2}$, which is close to what 
we find as well. They find no line of sight with column density smaller than 
$\simeq 1 \times 10^{20}$ cm$^{-2}$. In our case the smallest column density is about $7 \times 10^{19}$
cm$^{-2}$. Note that the smallest column density of atomic gas observed in the Galaxy is about 
$4.5 \times 10^{19}$ cm$^{-2}$ (Lockman et al. 1986).

Figure~\ref{rap_col_dens} shows the histogram of the fraction, $r$  of 
WNM (full lines), CNM (dotted lines), thermally unstable gas (dashed lines) along the lines 
of sight, e.g. $r _{\rm cnm} = n_{\rm c,cnm}/n_{\rm c, tot}$. 
Note the definition follows those of paper I, i.e. WNM is gas having temperature
higher than 5000 K whereas unstable gas has temperature between 200 and 5000 K.  
Top (Bottom) panel is for line of sight parallel to y-axis (x-axis).
The fraction of CNM can be compared with bottom panel of Fig.~7 from Heiles \& Troland (2003). 
Again both appear to be  similar. The median value in our case is about 0.34, whereas Heiles \& Troland quote
0.3. We have about 10$\%$ of lines of sight for which the fraction of CNM is smaller than 0.05 and they 
find about 30$\%$  of lines of sight with no CNM. We have about 30$\%$ of lines of sight with a fraction 
of CNM larger than 0.5, whereas they find this fraction to be about 10$\%$. 

Figure~\ref{fract_cnm} displays the fraction of CNM, $r _{\rm cnm}$, as a function of the 
total column density along the line of sight. It can be compared with Fig.~8 of 
Heiles \& Troland (2003). A similar trend is observed in both cases, $r _{\rm cnm}$
increases with the total column density. However, the observations appear to have a  
dispersion larger than the simulation results. This could possibly be a consequence of 
the fact that the observations probe several regions which are physically uncorrelated.

Altogether, these numbers appear to be similar. One should however keep in mind that our statistics 
are based on correlated lines of sight, whereas the lines of sight observed in the Millennium survey represent 
regions with a priori different physical conditions. As emphasized in paper I,  physical conditions
like external triggering and level of turbulence do influence the physical parameters of the flow as the 
CNM/WNM fractions.
Indeed, since the region that we are simulating is actively forming CNM structures, it is unsurprising
that the number of lines of sight with no CNM is underestimated and the fraction of lines of sight dominated
 by CNM is overestimated.
Nevertheless, the relatively good agreement  suggests  that the numerical setup we used leads to a 
reasonably realistic medium.

Another useful statistics is the number of lines of 
sight which cross a piece of fluid of a given density, $n$.
The result is shown on top panel of Fig.~\ref{line_sight} (logarithmic intervals 
of density are used). While most 
lines of sight intercept fluid particles of density between 0.5 and 100 
cm$^{-3}$, there is about 10$\%$ lines of sight which cross gas denser 
than 1000 cm$^{-3}$ and less than 0.1$\%$ which cross gas denser than 
10$^4$ cm$^{-3}$. In particular, these results suggest that the so-called TSAS 
(Heiles 1997) are somehow rare events as suggested by 
Stanimirov\'ic et al. (2003) and Johnstone et al. (2003), 
even if quantitative prediction should be considered 
with great care.

Finally, an interesting question is: when a line of sight crosses 
a fluid particle of a given density, what is the mean column density 
of gas at this density along the line of sight ? The result is shown 
in the second panel of Fig.~\ref{line_sight} (note that we have plotted the mean 
column density of gas at given density, $\rho_0$ along lines of sight which do 
intercept a fluid particle of density such that  
$log \rho_0$ and $log \rho_0 + d log\rho_0$). While the mean column 
densities of gas at density  1-2 cm$^{-3}$ and $\simeq$100 cm$^{-3}$ 
are approximately equal to about 3$\times 10^{19}$ cm$^{-2}$, the 
mean column densities of gas denser than $10^{3}$ cm$^{-3}$ are about 2 to 
3 times larger.  This means that, although few lines of sight intercept 
fluid particles denser than 10$^3$ cm$^{-3}$, when it is the case, 
the column density of the dense gas is usually a significant fraction of the 
total column density along the line of sight.

\section{HI spectra}
\label{spectra} 

\subsection{Emission and absorption}

\begin{figure*}
\includegraphics[width=16cm]{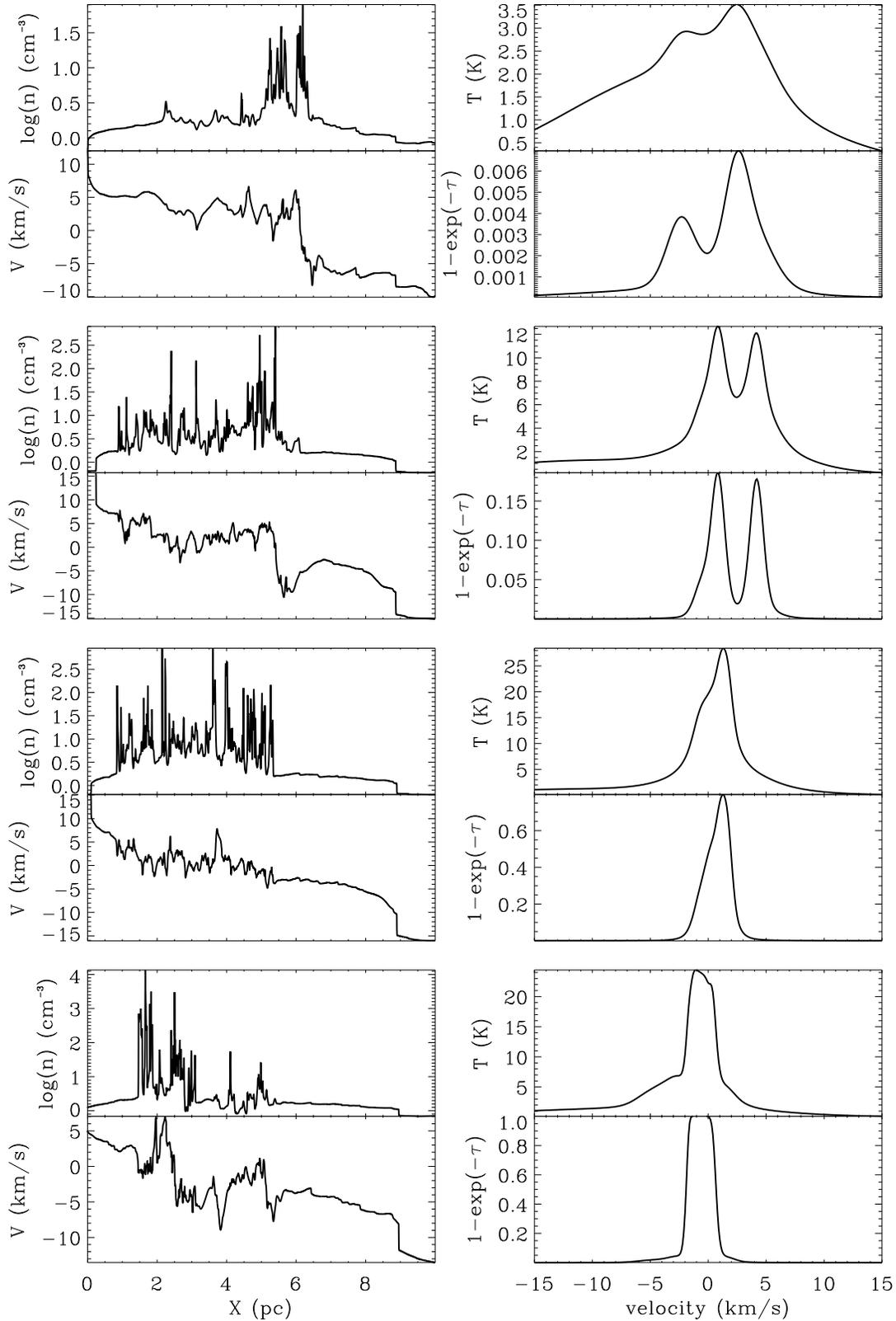}
\caption{Density and velocity fields along 4 lines of sight 
(parallel to x-axis) and 
 synthetic HI spectra in emission and in absorption calculated along 
these lines of sight.}
\label{coupe_spectre}
\end{figure*}

\begin{figure*}
\includegraphics[width=16cm]{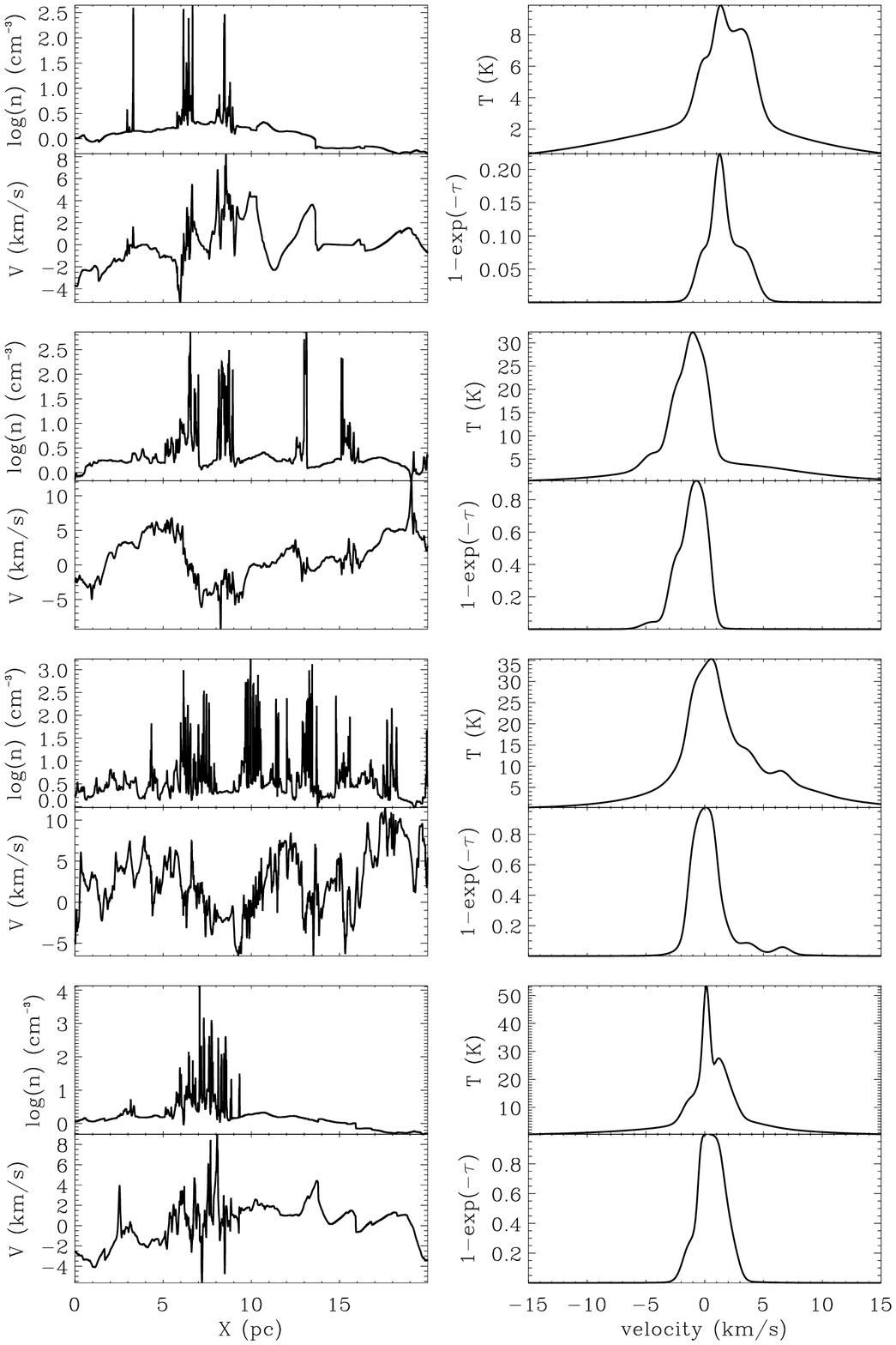}
\caption{Density and velocity fields along 4 lines of sight 
(parallel to y-axis) and 
 synthetic HI spectra in emission and in absorption calculated along 
these lines of sight.}
\label{coupe_spectre_y}
\end{figure*}

Since HI spectra provide most of the available information on the 
interstellar atomic hydrogen, we give here various 
examples of synthetic HI spectra calculated from the numerical simulation, 
in emission and in absorption.  This is certainly 
the most straightforward way of comparing observations and simulations. 
Figure~\ref{coupe_spectre} shows the density and the velocity 
fields along 4 lines of sight parallel to the x-axis 
(only half of the box simulation 
is displayed since the other half contains only WNM and no 
CNM structure) as well as the 21cm line emission and  extinction.

The first case (top panels) shows a line of sight with low density CNM structures 
($\simeq 30$ cm$^{-3}$). Two groups of clouds can be seen ($x \simeq 5.5$ and $\simeq 6$  pc), 
each of them being strongly substructured. The relative velocity 
of the two groups of clouds is about 5 km/s and each of them presents 
an internal velocity dispersion of a few km/s. In spite of 
this complexity, the emission spectrum appears to be relatively 
smooth. This is a consequence  of the fact that the thermal broadening of the 
HI 21cm line emission is not small  compared to the turbulent broadening.
The large component of width $\simeq$ 20 km/s is  the 
WNM emission. The two narrow components seen in emission and in absorption
are due to the contribution of the two groups of clouds. The largest 
extinction is about 0.007 and the width of the lines is typically
around 3-5 km/s. These numbers are very similar to the numbers quoted
 in Braun \& Kanekar (2005) and Stanimirovi\'c \& Heiles (2005).

The second case (second line of panels) shows a line of sight which contains 
four CNM structures of density higher than 100 cm$^{-3}$. Whereas three
of them ($x \simeq 2.5$, $x \simeq 3.2$ and $x \simeq 5.5$ pc) 
present a modest velocity dispersion of few km/s
 with respect to the others, one of them 
(located at $x \simeq 5$ pc) moves at  a velocity 
of about 4 km/s with respect to the other. 
As a consequence,  the corresponding HI spectra present 
two peaks, the first one ($v \simeq$ 1 km/s) 
being due to the contribution of the three structures 
moving at approximately the same velocities and the second ($v \simeq$ 4 km/s)
 being due to the fourth structure ($x \simeq 5$ pc). Whereas the first peak is slightly 
broadened by the velocity dispersion of the 3 structures, the width of 
the second one is mainly thermal.

The third case shows about 10 CNM structures denser than 100 cm$^{-3}$
and about the same number of CNM structures having a lower density.  
A velocity gradient of about 
1 km/s/pc takes place between $x \simeq 1$ and $x \simeq 6$ pc. 
As a consequence, the HI spectra present a single peak, instead of 2 in 
the preceding cases, which is  significantly 
broadened by the velocity dispersion of the structures. Note that this 
spectrum is qualitatively similar (temperature, extinction and shape) 
to the spectra shown by Heiles (2001) except
for the width which is 2 times larger than the case presented here.

The fourth case shows a line of sight which crosses the highest 
density reached in Fig.~1 of paper II 
($x \simeq 1.65$ pc and $y \simeq 7.07$pc). The HI 21cm line is strongly 
saturated because of the large column density. The spectra are 
again broadened by the velocity dispersion between CNM structures. Again, 
  the complex structure of the flow cannot be easily recovered from the 
HI spectra which are relatively smooth.

Figure~\ref{coupe_spectre_y} shows four lines of sight 
along the y-axis. Indeed, since our experimental setup is strongly 
anisotropic, it is worth to see what are the consequences of this anisotropy on the 
synthetic HI spectra. In particular, along the y-axis few uncorrelated groups
of structures can contribute to the synthetic signals. 
This is well illustrated by the first, second and third cases in which 
3 to 5 groups of structures  are present. Each group presents an internal 
velocity dispersion of a few km/s and an average  velocity with respect to the 
other groups which  can be as high as $\simeq 5$ km/s. Interestingly, 
this is somehow
similar to the structures obtained by Hennebelle \& Passot (2006) in the case
where significant Alfv\'en waves are initially present in the flow. 
Whereas in their case, this was due to the effect of Alfv\'en waves 
which tend to subfragment the CNM structures,  here the 
subfragmentation of the group of structures is likely due to 
the turbulent fluctuations. As a consequence, the synthetic spectra are 
on average broader than the spectra obtained along the x-axis (factor 
$\simeq 1.3-1.5$). 

The fourth case  (Fig.~\ref{coupe_spectre_y}) corresponds to the 
line of sight parallel to the y-axis which crosses the highest 
density reached in the simulation. As for the fourth case
 (Fig.~\ref{coupe_spectre}), the 21cm line is strongly 
saturated. The narrow peak seen in emission 
is produced by the strong density fluctuation (located at $x \simeq 7$ pc)
which is also displayed in Fig.~5 of paper II. As explained 
in paper II, we believe that this structure is a good candidate to 
explain the TSAS observed in the atomic gas. This spectrum  could
therefore constitute a typical signature of such event. 
Note that in the fourth 
case of Fig.~\ref{coupe_spectre}, this narrow component is not seen because 
in front of it, stands a less dense structure ($\simeq 10^3$ cm$^{-3}$)
which has a high extinction. On the contrary,  in the present case, there is 
no dense structure on the right side of the densest peak.

\subsection{Effect of finite resolution}

\begin{figure*}
\includegraphics[width=16cm]{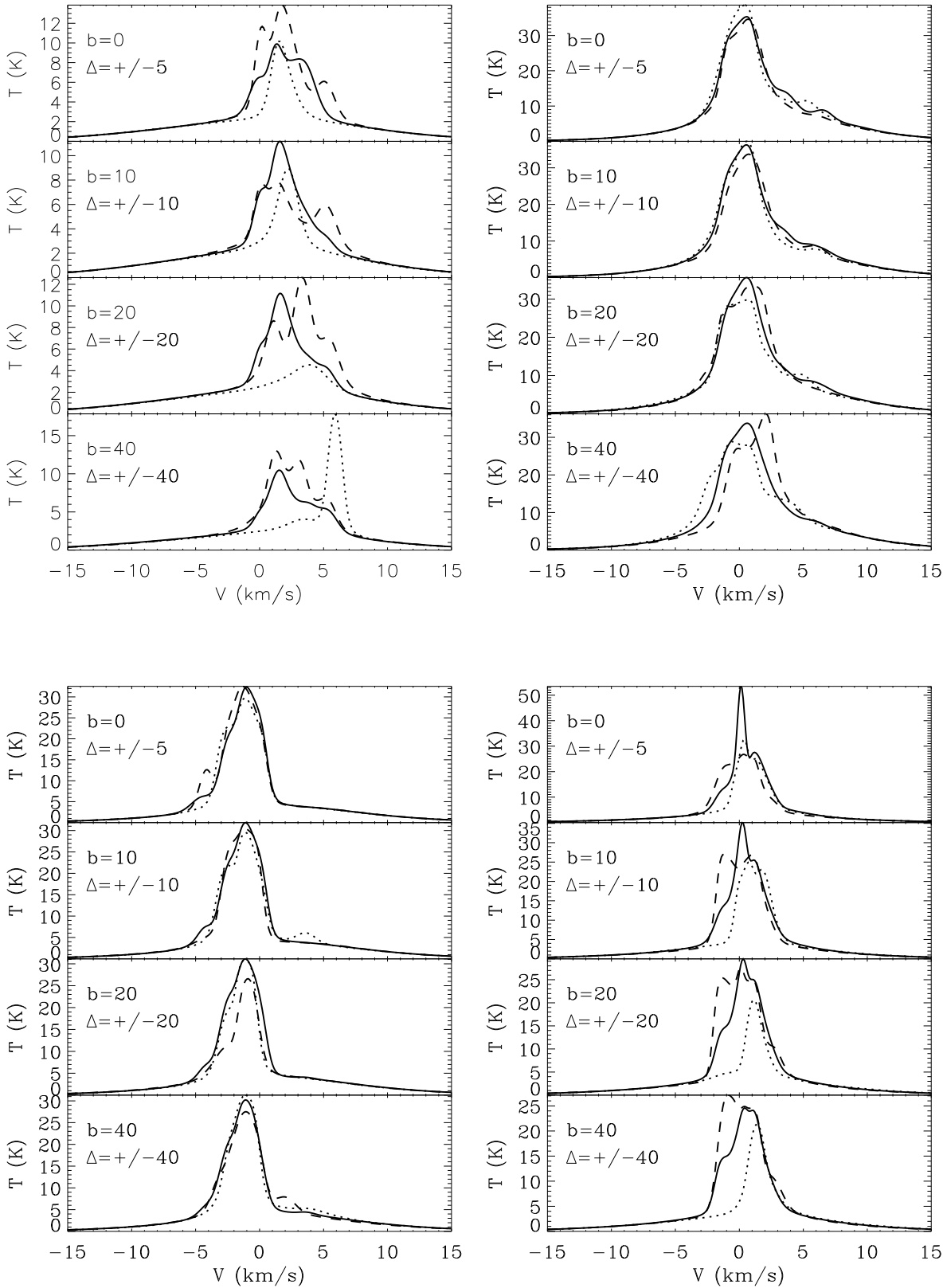}
\caption{Four series of spectra showing the influence of a finite 
observational spatial resolution. 
The full line of first panel of each series show the same spectra than those 
shown in Fig.~\ref{coupe_spectre_y}, the full lines of second, third and fourth
panels show the spectra obtained by convolving the spectrum distribution by a
Gaussian of width at half maximum equal to respectively $b=10$, 20 and 40 cells.
The  dotted (dashed) lines of each panel show the unsmoothed spectra shifted from the central 
position by   $\Delta=$-5 (5), -10 (10), -20 (20) and -40 (40) cells respectively.  }
\label{spectre_smooth}
\end{figure*}

Here, we investigate the effect of finite resolution on the HI emission spectra. 
We convolve the HI spectra calculated along the lines 
of sight by  Gaussian functions of various widths at half maximum,  $b$. 
 Figure~\ref{spectre_smooth} shows the results. 
Four series of spectra showing the influence of a finite resolution are displayed. 
The full line of first panel of each series shows the same spectra as those 
shown in Fig.~\ref{coupe_spectre_y}, (left-top series of spectra is first case of 
Fig.~\ref{coupe_spectre_y}, left-bottom is second case, right-top is third case, whereas
 left-bottom is fourth case).
The full line of second, third and fourth
panels of each series shows the spectra obtained by convolving the spectrum distribution by a
Gaussian function of width at half maximum equal to  $b=10$, 20 and 40 respectively.
The dotted (dashed) lines of each panel show the unsmoothed spectra shifted from the central 
position by   $\Delta=$-5 (5), -10 (10), -20 (20) and -40 (40) cells respectively.

Left-top series (first case of Fig.~\ref{coupe_spectre_y})  show that there is 
a significant variability of the HI 21cm emission spectra when different lines of sight 
are selected (dotted and dashed lines). This has severe consequences on the convolved spectra which 
vary significantly when the width at half maximum of the Gaussian function used for the convolution,
 increases. As can be seen, the shape of the unconvolved spectra (first panel) 
  is rather different from the shape obtained with $b=10$, 20 and 40. 

Left-bottom series (second case of Fig.~\ref{coupe_spectre_y}) and 
right-top series (third case  of Fig.~\ref{coupe_spectre_y}) show cases
 for which the variability
of the spectra is much less important than in the previous case. The general shape of 
the convolved spectra (full lines) does not change much, the fluctuations appear to  
be on average smaller than $\simeq 10 \% $ with some stronger fluctuations 
of about $\simeq 20-30 \%$. The  difference between this behaviour and what 
has been found in the previous case, is largely due to the fact that the CNM structures
are much smaller in the latter than in the former. Therefore the spatial length 
over which the column density varies significantly, is also smaller in left-top case than 
in these 2 cases, making the dependency on the lobe effect more important.

Right-bottom series (fourth case of Fig.~\ref{coupe_spectre_y}) show
also significant variability in spite of the large emission value. This is due to
the fact that the line of sight ($\Delta=0$) crosses a small scale and dense structure 
(seen in Fig.~5 of paper II)
which results from a strong collision between 2 CNM structures. Therefore, as indicated in 
Fig.~\ref{line_sight}, the column density varies significantly along the line of sight in 
few cells. 

On average,  the smoothed spectra are not much broader than the unsmoothed one. This is certainly a 
consequence of the fact that the velocity dispersion along the lines of sight presents variations 
which are smaller than its average value (Sect.~2.1).

\section{Clouds properties}
\label{cloud}

In paper II, the mass spectrum of the structures formed in the 
simulation has been measured and theoretical arguments to explain it,
have been presented. Here, we further study the structures properties.  
First, we consider the case of structures simply defined 
by the clipping procedure explained in paper I and II. Second in order
to quantify the fact that the structures are spatially correlated and 
not randomly distributed, we investigate some properties of 
 groups of structures.

\subsection{Properties of individual clouds}
We study various structure  properties which could 
be compared with observations or with other theoretical works. 
We pay special attention to the distribution of column density 
within clouds, since an interesting observational result has been obtained by 
Heiles \& Troland (2005) with which comparison is possible.

In the following, we define the cloud size, $L$, as 
$L = \sqrt{I_1/M}$, where $M$ is the cloud mass whereas 
$I_1$ is the highest inertial momentum ($I_2$ being the smallest), 
i.e. the highest eigenvalue
 of the inertial matrix, ${\cal I}$, defined as ${\cal I}_{xx} = \int  y^2 dm$, 
${\cal I}_{yy}= \int  x^2 dm$ and 
${\cal I}_{xy}= {\cal I}_{yx} = - \int  x y dm$.

\subsubsection{Mass-size  relation and size distribution} 
\begin{figure}
\includegraphics[width=8cm]{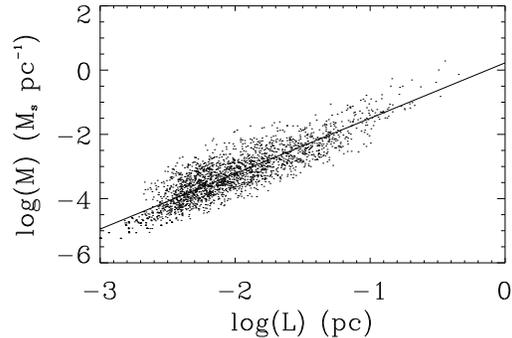}
\caption{Mass of the structures as a function of their size.
One finds the relation $M \propto L^\gamma$ with 
$\gamma \simeq 1.7$.}
\label{mass_L}
\end{figure}

%\begin{figure}
%\includegraphics[width=8cm]{densmoy_L.ps}
%\caption{Median density within one structure as a function 
% its size. A weak correlation is seen leading to 
%$n_{med}= L^\lambda$ with $\lambda \simeq 0.2$, although there is a
%considerable dispersion.}
%\label{densmoy_L}
%\end{figure}

An interesting property for cloud characterization is the mass-size 
relation. 
Figure~\ref{mass_L} shows the mass of the structure as a function of their size. 
A clear correlation is seen, leading to the relation 
$M \propto L^\gamma$ with $\gamma \simeq 1.7$. 

Since, as will be seen below, this relation turns out to be useful to understand 
the column density distribution, it is worth to understand its physical origin. 
For this purpose,
we consider the distribution function, ${\cal N}(L)$,  of structures of size $L$.
The number of structures of size between $L$ and $L+dL$ 
is equal to the number of structures of mass between  $M$ and $M+dM$, 
  ${\cal N}(L) dL = {\cal N}(M) dM$ where $M \propto L^\gamma$. 
In paper II, we obtain that ${\cal N}(M) \propto M^{-\beta}$ with $\beta \simeq 1.7$. 
Therefore, we have: ${\cal N}(L) = {\cal N}(M) \, dM/dL \propto L ^{-\beta  \gamma + \gamma -1}$.
With the values of $\beta$ and $\gamma$ given above, we have:
$-\beta  \gamma + \gamma -1 \simeq -2.19 $. This implies that 
${\cal N}(L) L^2 \simeq L^{-0.19}$, i.e. the number of structures of a given scale $L$ within 
a (bidimensional) volume $L^2$, is almost the same as the number of structures of 
 scale $L'$ within a volume $L'^2$. This implies that the flow is nearly scale-invariant 
(see Elmegreen 1997 and Padoan \& Nordlund 2002). This
 is likely a consequence of the 
flow being turbulent, although in the present problem, there are characteristic scales
which may explain the deviation from the scale-invariant behaviour. Another possibility is
that the size of the simulation is still insufficient and the numerical mesh, as well as the
injection scale, still have  an influence.

\subsubsection{Cloud shape}
\begin{figure}
\includegraphics[width=8cm]{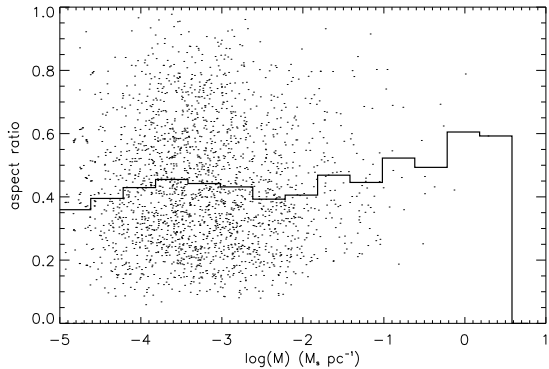}
\caption{Aspect ratio of the clouds as a function of their mass. 
The solid line shows the average value per logarithmic interval of mass.}
\label{rap_asp}
\end{figure}

\begin{figure}
\includegraphics[width=8cm]{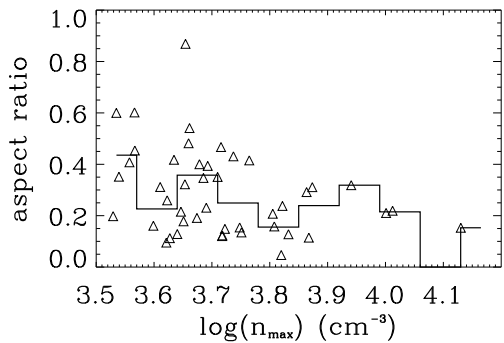}
\caption{Aspect ratio of the densest regions   as a function of the 
largest gas density within the region. 
The solid line shows the average value per density interval.}
\label{rap_asp_tsas}
\end{figure}

Figure~\ref{rap_asp} shows the aspect ratio of the cloud defined
as $\sqrt{I_2/I_1}$. The solid line represents the mean value per logarithmic  
 mass interval.  The mean aspect ratio is approximately constant and equal to 
about 0.4 until $M \simeq 0.1$ M$_s$/pc. It then increases to about 0.6 for larger mass.
Nevertheless we note  that there is a considerable dispersion, some structures being 
very elongated.

Another related question is the shape of the very dense regions observed in the simulations
(see section 3.2 of paper II). Since this high density component is created by shocks, it is 
expected that these regions should be very elongated. Figure~\ref{rap_asp_tsas} shows 
the aspect ratio of these regions as a function of the peak density (note that since these regions
are rare, 6 time steps have been used to provide this plot) and confirms that these structures
are elongated with a mean aspect ratio of about $\simeq 0.3$. We note that a significant fraction 
of them has an aspect ratio smaller than 0.2, which is significantly smaller than the mean 
aspect ratio of unshocked CNM structures. Indeed very few CNM clouds have an aspect ratio smaller 
than 0.2.  This suggests that this could constitute an interesting test for the assumption of 
TSAS being shocked CNM structures.

%\subsubsection{Fractal index} 

%\begin{figure}
%\includegraphics[width=8cm]{fract.ps}
%\caption{Measure of the fractal index defined 
%as the index, $f$, in the relation $S = P^f$ between 
%the surface, $S$ and the perimeter, $P$. One finds
%$f \simeq 1.2$.   }
%\label{fract_index}
%\end{figure}

%We now determine the fractal index, $f$, of the CNM structures. 
%This is indeed an interesting way of characterizing the structures further. 
%This could also be used to compare with observations as well as with 
%other theoretical works. 
%In order to do this, we simply compute the surface and the perimeter of the 
%structures and plot the logarithm of these two quantities. 
%The result is shown in Fig.~\ref{fill_fact}.
%We find that $S \simeq p^f$ where $f \simeq 1.2$.
%This value is consistent with the CNM structures being 
%closer to be linear objects than 2D ones and is good agreement
%with the filamentary structures which are seen in Fig.~2 of paper II
%and the aspect ratio measured in Fig.~\ref{rap_asp}. 

\subsection{Column density distribution of  clouds}
In the quest for comparison between observations and simulations, an
 interesting result obtained by Heiles \& Troland (2005), appears 
to  deserve special attention.
Heiles \& Troland show that the column density distribution of 
their CNM structures follows the relation ${\cal N}(N) \propto 1/N$.
In this section, we study in some details the  column 
density distribution of the CNM structures and propose a theoretical explanation for 
this relation. 

\subsubsection{Numerical results}
\begin{figure}
\includegraphics[width=9cm,angle=0]{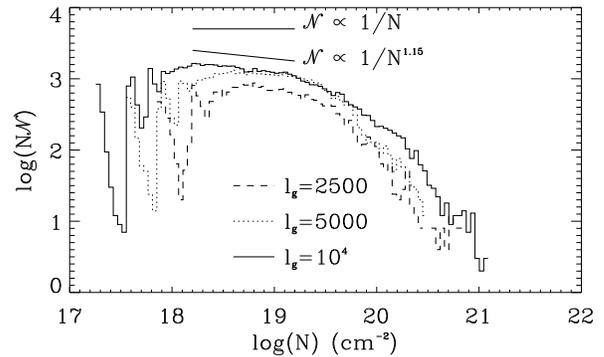}
\caption{Number of lines of sight (crossing a single CNM structure) 
per logarithmic   column density interval.
 Full line corresponds to the $(10^4)^2$ cells simulation, dotted line
is for the $(5000)^2$ one,  dashed line is for 
the $(2500)^2$ cells simulation.}
\label{column_density}
\end{figure}

Here, we display  the  column density distribution, ${\cal N}(N)$ obtained 
by putting together the column density distribution of all the structures
identified in our simulation.
The result is shown in Fig.~\ref{column_density} 
which shows the number of lines of sight, $n_l$,  per logarithmic interval 
of column density, $d n_l / d{\rm log} N = N d n_l / d N = N {\cal N}(N)$, for various
numerical resolutions. 

The low column density part is obviously spoiled by numerical resolution 
effects. In the intermediate part, say between $3 \times 10^{18}$ cm$^{-2}$ 
 and $\simeq 3 \times 10^{19}$cm$^{-2}$, 
the distribution does not appear to depend on the numerical resolution for $l_g > 5000$.
This suggests that in this range of column density, numerical convergence has been 
reached. On the contrary, the high column density part strongly depends on
the numerical resolution. In particular the higher the resolution, the larger the 
number of lines of sight having high column density. We therefore believe that only the intermediate 
column density part is physically significant. 

As can be seen from the highest resolution simulation (solid line) displayed in 
Fig.~\ref{column_density},  the column density distribution 
is stiffer between $1.5 \times 10^{19}$ and $3 \times 10^{19}$ cm$^{-2}$ than  between 
$1.5 \times 10^{18}$ and $1.5 \times 10^{19}$ cm$^{-2}$. Since there is no 
clear physical reason for this to be true, this is most likely a numerical effect due to the 
finite size of our numerical experiment. 
Since the size of the simulation box
is only 20 pc and the WNM density about 0.8 cm$^{-3}$, it is indeed difficult to produce 
structures with column density equal to, or higher, than 
about $20 \times 3.08 \times 10^{18} \times 0.8 \simeq 5 \times 10^{19}$ cm$^{-2}$.
Therefore one expects that the number of big structures and therefore the number of 
lines of sight with large column density drop above a certain threshold. The value of
this threshold for the column density is certainly smaller than  
$\simeq 5 \times 10 ^{19}$ cm$^{-2}$. This is in good agreement with what has been found 
in paper II for the mass spectrum (Figure 15) which presents a cutoff for high mass as well.

Therefore, we believe that the most reliable part of the column density distribution 
is the one between  $\simeq 1.5 \times 10^{18}$ and $1.5 \times 10^{19}$ cm$^{-2}$. Note that 
this range of column density corresponds to the part of the mass distribution, 
(Fig. 15-18 of paper II) for which numerical convergence has been reached.
The solid line shows that between $1.5 \times 10^{18}$ and $1.5 \times 10^{19}$ cm$^{-2}$, 
 one has about ${\cal N}(N) \propto 1/N^{1.15} \simeq 1/N^{1.2}$
which is slightly stiffer but nevertheless close to the 
${\cal N}(N) \propto 1/N$ distribution. Although the disagreement appears to be 
statistically significant (Heiles 2006, Private Communication), 
it appears to be  mainly due to the  column densities
larger than $10^{20}$ cm$^{-2}$, which we cannot probe here. It is also possible that the  column
 density distribution is biased because of large column density components being blends of two smaller 
components or because of small column density being missed (Heiles 2006, Private Communication).

Note that Heiles \& Troland (2005)   infer
${\cal N}(N) \propto 1/N$  for $N$ between $2 \times 10^{18}$ and $2 \times 10^{20}$ cm$^{-2}$.
The reason of this disagreement is probably due, at least in part, to the numerical setup
and most likely to the finite spatial range of our simulation box. Future studies 
using larger box size and larger numerical resolution, may give answer to these questions.

\subsubsection{Physical explanation}
\begin{figure}
\includegraphics[width=8cm]{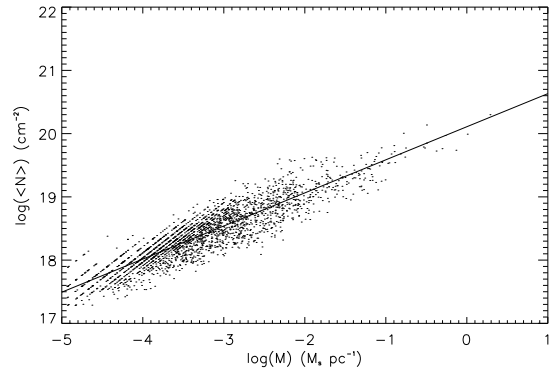}
\caption{Average column density of the structures as a function of their mass.
One finds the relation $<N> \propto M^\eta$ with 
$\eta \simeq 0.52$.}
\label{col_dens_mass}
\end{figure}

Here, we propose a physical explanation for the  column density distribution,
 ${\cal N}(N) \propto N ^{-\alpha}$.

For this purpose, we make the assumption that the clouds are sufficiently 
uniform for us to consider that the column density within the cloud does not 
strongly fluctuate.  Figure~\ref{col_dens_mass} 
displays the mean structure column density as a function of their mass, $M$. 
A very good agreement is obtained with the relation $<N> \propto M^{0.5}$.

This result can be understood as follows. 
Let us consider a spherical CNM cloud of uniform density  $n$. Its mass can be written 
$M \simeq n L^2$ and the column density $N \simeq n L$. Thus
$<N> \propto M ^{\eta}$ with $\eta = 0.5$. In the case we are considering,
$M \propto L^\gamma$ which implies that the CNM structures are not 
spherical clouds with uniform density. To fix idea, we consider 2 cases.

First, the CNM clouds are roughly roundish  but the density is not uniform 
because the internal structure is complex.
The number of lines of sight crossing the structure 
is proportional to $L$, thus the mean column density follows
$<N> \propto L ^{\gamma -1} \propto M ^{(\gamma -1)/\gamma}$. 
With $\gamma = 1.7$, we have $(\gamma-1)/ \gamma \simeq 0.41$.

Second, the CNM clouds are filaments with uniform density. The major
axis has a length $L$, whereas the minor axis has a length $L^{\gamma -1}$.
If the filament is seen along the minor axis, the mean column 
density again follows  $<N> \propto L ^{\gamma -1} \propto M ^{(\gamma -1)/\gamma}$. 
If the filament is seen along the major axis, then 
$<N> \propto L  \propto M ^{1/\gamma}$ with $1/\gamma \simeq 0.59$.  

Since all the values found are slightly above or slightly below 0.5, it is 
not surprising to find $\eta \simeq 0.5$. \\

The number of lines of sight crossing a structure of column density
between $N$ and $N +d N$ is proportional to the number of structures having a 
mass between $M$ and $M+dM$ times the length of the structures (since the number of line
of sight crossing a structure of size $L$ is proportional to $L$). Thus we have:
${\cal N}(N) dN \propto {\cal N}(M) \times M^{1/\gamma} \times dM$. 
This leads to ${\cal N}(N) \simeq N ^{(-\beta + 1/\gamma + 1-\eta) / \eta}$.
Since with $\beta$=1.7, $\gamma$=1.7 and $\eta$=0.5 we have $-(-\beta + 1/\gamma + 1-\eta) / \eta \simeq 1.22$, 
we get: ${\cal N}(N) \propto 1/N^{1.22}$. This value is remarkably close to what is measured in 
the simulation. 

The good agreement between the analytical value and the exponent measured
in the simulation, suggests the validity of the approach. However, 
since we used the value $\gamma$, inferred from the simulation, it could be 
that its value is different in the ISM (as  previously discussed, the size of 
the computational 
box or the numerical resolution could be  insufficient). In particular, it could be, 
that real atomic flows are closer from strict scale invariance than what 
has been found in the simulation. 
If instead of taking 
the value of $\gamma$ measured in the simulation, we assume ${\cal N}(L) \propto L^{-2}$,
 thus $L^{-2} dL = M^{-\beta} dM$ 
which leads to $M \propto L^\gamma$, $\gamma = 1/ (\beta  - 1)$. 
With this value of $\gamma$, we have $-(-\beta + 1/\gamma + 1-\eta) / \eta = 1$ 
which is exactly the value inferred by  Heiles \& Troland (2005). 
Interestingly, the result appears to be independent of $\beta$ and $\eta$.

Let us stress that the result, ${\cal N}(N) \propto N ^ {-\alpha}$, with 
$\alpha \simeq 1-1.2$,  is largely  a consequence  of ${\cal N}(L) \propto L^{-2}$ and of the fact that 
the column density across a structure is nearly proportional to its size, which is due to 
the thermally bistable nature of the flow, since in that case, the density 
is determined almost entirely by the mean pressure and does not depend on 
scale or mass. 
Future works, using  larger simulations, may clarify 
to which extent such flow follows  ${\cal N}(L) \propto L^{-2}$.

\subsubsection{The 3D case}
Since the simulation and the theoretical derivation are bidimensional, 
it is worth investigating the value of the exponent of the column density distribution
in the 3D case.
The theory developed in paper II for the structure mass spectrum, ${\cal N}(M)$, predicts that 
${\cal N}(M) \propto M ^ {-\beta}$ with $\beta \simeq 16/9$. 
Assuming scale invariance, we get ${\cal N}(L) \propto 1/ L^3 $, which leads to:
$L^{-3} dL = M^{-\beta} dM$ and thus to $M \propto L^\gamma$, with $\gamma=2/(\beta -1)$.
We again make the assumption that  $<N> \propto M^\eta$. For a spheroidal cloud with 
uniform density, we have obviously  $\eta =1/3$. In 3D, we have 
 ${\cal N}(N) dN = {\cal N}(M) \times L^2 \times dM$. Thus we  obtain
${\cal N}(N) \simeq N ^{(-\beta + 2/\gamma + 1-\eta) / \eta} = N ^{-1}$, 
  which is again  the value  inferred by 
Heiles \& Troland (2005) (as in the 2D case, it does not depend  on $\beta$ or on $\eta$). 

Note that for $\beta=1.8$, we get $\gamma = 2/ 0.8 = 2.5$, implying 
$M \propto L^{2.5}$.  This is very similar for the mass-size relation inferred 
by Larson (1981) for the molecular clouds. 
Even more interesting is the comparison with the results obtained by Heithausen et al. 
(1998) for CO clumps of mass ranging from $10^4$ solar mass down to Jupiter mass. 
For these clouds, they measure $M \propto L ^{2.3}$, ${\cal N}(M) \propto M ^{-1.8}$
and ${\cal N}(L) \propto L ^{-3}$. This is  consistent with what we  
infer for the CNM  clouds.

\subsection{Group of structures properties}
\label{group}
\begin{figure*}
\includegraphics[width=15cm,angle=90]{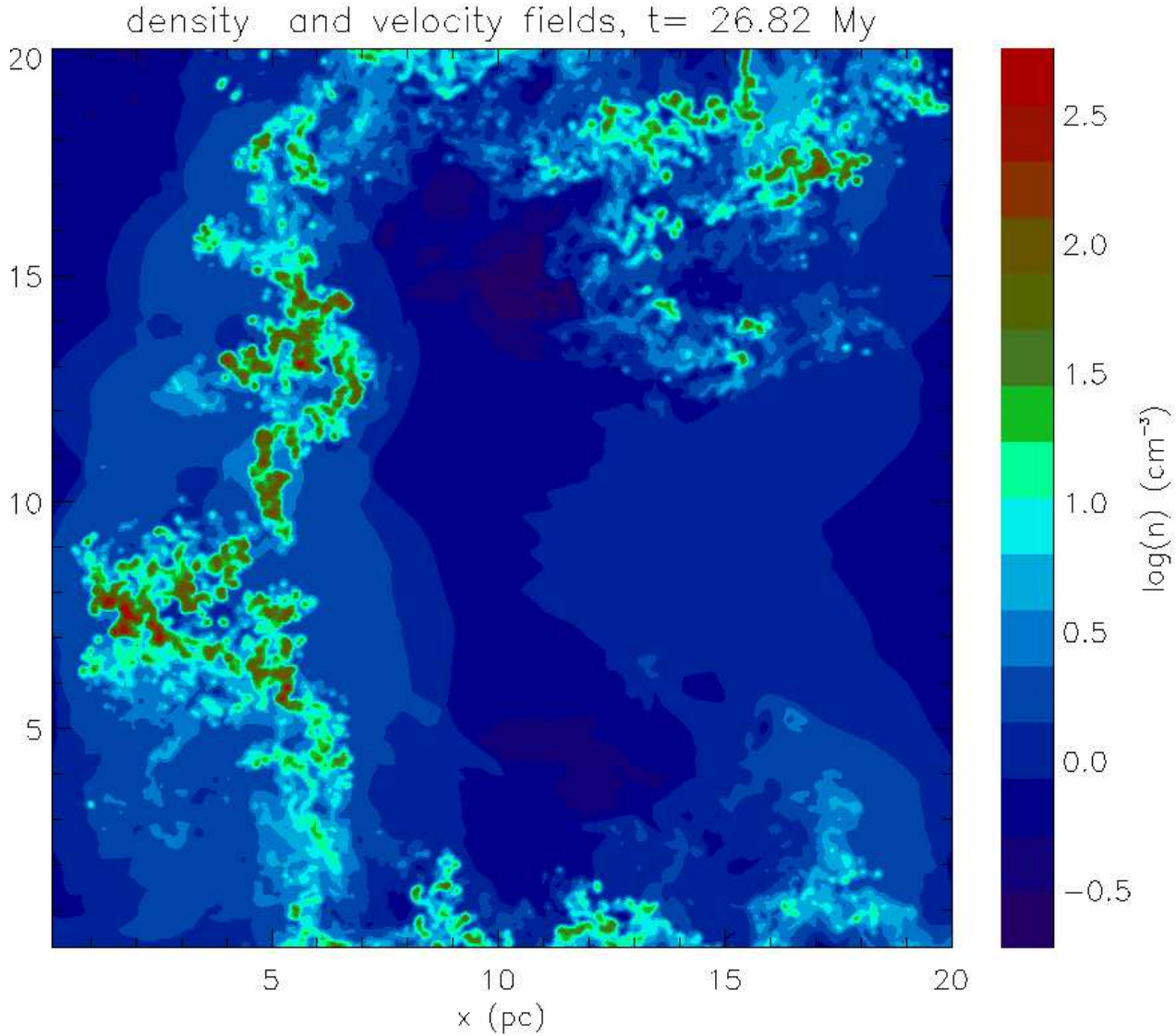}
\caption{Smoothed density field. The original density field 
has been convolved with a Gaussian having a width at half maximum 
 equal to $b=40$ cells leading to a spatial resolution of about 
0.1 pc. }
\label{smooth_density}
\end{figure*}

The aim of this section is to  further characterize the structure of the flow and 
in particular the fact that the CNM structures appear to be highly correlated and 
not randomly distributed.
\subsubsection{Methods}
 In order to achieve this, we smooth the density field 
by convolving it with  Gaussian  of various widths at half maximum, $b$. 
This has the advantage 
of mimicking observations done with a telescope having a gaussian lobe and a 
 spatial resolution equal to $b \times dx$, where $dx$ is the
size of one of our cells. For $b=40$, this leads to a resolution of about 0.08 pc. At 100 pc
of distance, this corresponds to a resolution of 3 arc minutes which is 
comparable to the resolution of the Arecibo telescope.  
 Figure~\ref{smooth_density} shows the 
 density field displayed in Fig.~1 of paper II  which has been 
convolved by a Gaussian of width at half maximum equal to $b=40$ cells. 
As expected, the field displayed in Fig.~\ref{smooth_density} is  more uniform than 
the field displayed in Fig.~1 of paper II, although significant large scale 
fluctuations  persist. As in paper II, we identify the structures by applying a 
simple clipping algorithm, i.e. the structures are defined as groups of connex cells denser
than a density threshold $\rho_0$. Two values of $\rho_0$ are considered in the following, 
namely $5$ and 30 cm$^{-3}$.

We stress that the analysis presented in this section, 
ought to characterize an  {\it ensemble} of individual, disconnected from 
each other, CNM structures which are  spatially correlated and would 
be seen with a telescope of insufficient spatial resolution as a coherent 
cloud.

\subsubsection{Velocity dispersion}

\begin{figure}
\includegraphics[width=8cm]{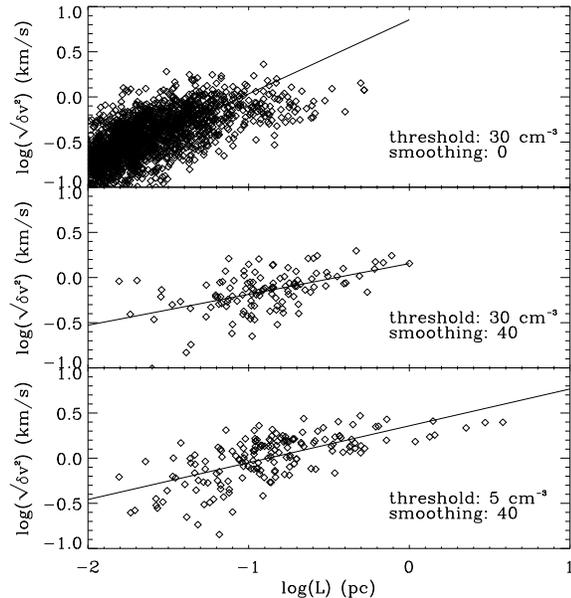}
\caption{Measure of the internal velocity dispersion of structures 
extracted from numerical data obtained with different smoothing and clipping 
threshold.}
\label{disp}
\end{figure}

The cloud velocity dispersion is an interesting  parameter 
for the understanding and the characterization of the turbulence which 
takes place within the atomic medium. Before we pursue, we would like 
to stress that the use of the word {\it turbulence} in referring to the gas motions should 
be made with some care. Indeed,  the structure of the flow studied here is 
very different from the incompressible fluids which have been extensively
studied and which constitute our basic conception of the turbulent processes.
As discussed above, the turbulence is produced by individual 
cloudlets motion rather than by eddies in a nearly homogeneous medium. 
In particular, the coupling between the various spatial scales is presumably 
strongly modified by the stiff density fronts.

The cloud velocity dispersion is defined as stated by Eq.~\ref{velo_disp_line}, 
where $v_0$ is the mean cloud velocity and where all cells 
having a density above the threshold and lying in the area defined 
by the smoothed density field have been counted (we stress that we use unsmoothed data 
to compute the velocity dispersion). 
Figure~\ref{disp} shows results for $b=0$ and 40 and clipping thresholds 
5 and 30 cm$^{-3}$. 

In the case with no smoothing, we find that 
most of the structures have a velocity dispersion lower than 
1 km/s meaning that most of structures have a subsonic velocity dispersion.
By performing a simple linear fit, we infer: 
$\sqrt{\delta v ^2} \simeq 7$ km/s $\times (r/1 {\rm pc})^{0.8}$. Note that 
the size of structures found with no smoothing is generally smaller than 0.1 pc
so that this relation is valid only up to $\sqrt{\delta v ^2} \la 1$ km/s.

With $b=40$, the structures are bigger and the velocity dispersion higher. 
As can be seen, it is not strongly affected by the clipping threshold 
(factor $\simeq$2).  Typical velocity dispersions are now around 1 km/s 
and up to 3 km/s for the biggest structures. 
We infer $\sqrt{\delta v ^2} \simeq 1.4 $ km/s $ \times (r/1 {\rm pc})^{0.34}$ 
for a clipping threshold of 30 cm$^{-3}$ and 
$\sqrt{\delta v ^2} \simeq 2.2 $ km/s $ \times (r/1 {\rm pc})^{0.4}$ 
for a clipping threshold of 5 cm$^{-3}$.  
We note that these 
values are reminiscent of the values observed in HI clouds
by Crovisier et al. (1981) and Heiles \& Troland (2003, 2005), 
although Figure~12 of Heiles \& Troland (2003) suggests that 
the mean value of the Mach number inside structures is about 3. 
In our case, it is slightly lower, 1.5-2.5, depending on the threshold which 
is used to define the structures. 
At this stage, it  seems difficult to do a more quantitative comparison
since the present simulations are only 2D. In particular, the  approach
used in this section,  excludes any projection effect along the line of sight.  

The fact that the internal velocity dispersion of CNM structures is low ($<$ 1 km/s)
but that the velocity dispersion of the groups of structures is comparable
to observations, suggests that the turbulence inside CNM clouds is
largely due to  individual long-living cloudlet motions.

Again, we note that the velocity dispersion-size relation is reminiscent of 
what is observed in the case of molecular clouds (Larson 1981). 

\subsubsection{Filling factor} 

\begin{figure}
\includegraphics[width=8cm]{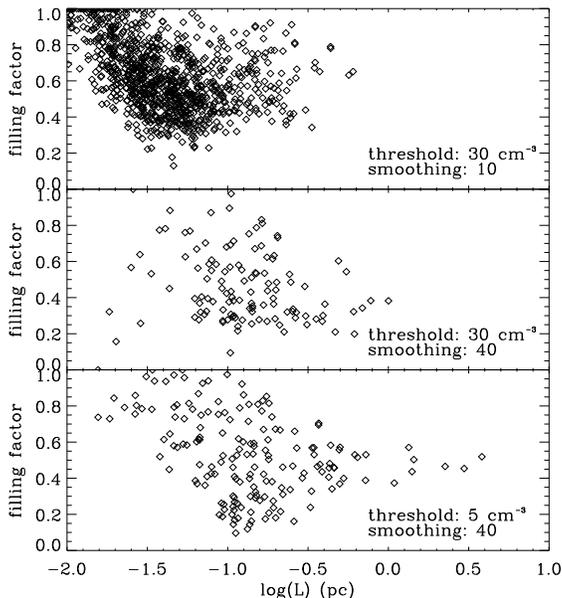}
\caption{Measure of the cloud filling factor in structures 
extracted from numerical data obtained with different smoothing and clipping 
threshold.}
\label{fill_fact}
\end{figure}

One of the striking aspects of the structures displayed in 
Fig.~1 and 2 of paper II is that the CNM appears to be 
very fragmented even in the densest parts of the flow. In order to quantify 
this effect, we have computed the filling factor of the structures 
extracted from the smoothed data. This is obtained by simply computing within 
these structures, the filling factor of the gas (taken from  unsmoothed data) 
having a density above the  clipping threshold.
The result is displayed in Fig.~\ref{fill_fact} which shows the 
filling factor for various $b$ and clipping thresholds. 
As expected, the value of the filling factor presents a significant scatter and 
depends on $b$, the width at half maximum of the Gaussian used to compute 
the smoothing. 
Typical values range nevertheless between 0.8 and 0.2 whereas
the average value is about $\simeq 0.5$.

\section{Conclusion}
In a companion paper, high resolution numerical simulations of the 
 turbulent interstellar atomic hydrogen have been presented. 
Because of the high spatial resolution (0.002 pc) reached in these
simulations, scales approaching the size of the smallest structures observed  in HI
can be described.

In this paper we have further analyzed these numerical data in order to make 
preliminary comparisons with HI observations. We provide statistical informations
on column density, average velocity and velocity fluctuations along the line of sight
and, when possible, we compare with the results  of the Millennium Arecibo 
21 centimeter absorption line carried out by Heiles \& Troland (2003, 2005).

We compute  HI 21cm line spectra in emission and in absorption along eight lines 
of sight. We reach the  conclusion that although these spectra are relatively
smoothed the lines of sight from which they are calculated appear to be highly complex
and fragmented. In particular, the "turbulence" is produced by the individual motions 
of few long living CNM structures rather than by motions within nearly homogeneous or 
isothermal  medium. 
This suggests that the Mach number deduced from observation is more indicative of 
the velocity dispersion from cloud to  cloud  rather than 
of the internal velocity dispersion of the clouds. 
We also consider the influence 
of a Gaussian lobe on the HI spectra and find that, it has in some cases, a drastic influence. 

To  further characterize the CNM structures produced in the simulation, 
we have studied some of their properties, namely
mass-size relation and shape. In an attempt to describe the high level of 
spatial correlations between the CNM 
structures and to mimic observations done with a limited spatial resolution, we have 
smoothed the data simulations by convolving them with a Gaussian function of various 
sizes at half maximum and analyzed the results. By doing this, it is possible to
 identify groups of structures that would be seen as a single object and study their 
properties like filling factor and velocity dispersion.
We find the mass-size relation, in the bidimensional simulation, $M \propto L^\gamma$ with
$\gamma \simeq 1.7$ and based on the assumption ${\cal N}(L) \propto L^{-3}$, 
we speculate that 
in 3D, we may have $M \propto L^{2.5}$. We also find that the velocity dispersion 
increases with the size of the structures, $L$, as  $\sqrt{<\delta v^2>} \propto L^{0.4}$.
These behaviour are reminiscent of the Larson laws's (Larson 1981, Heithausen et al. 1998)
 inferred for the molecular clouds. 
Moreover, since the mass spectrum that we obtain for the CNM structures in paper II, turns out 
to be compatible with the mass spectrum observed for the molecular clouds 
(Kramer et al. 1998, Heithausen et al. 1998), this suggests that (or at least is compatible with it), 
the origin of the molecular clouds, is determined at a very early stage in the diffuse atomic
medium. This is also consistent with the  proposition made by Falgarone \& Puget (1986)
 and recently reinvestigated by Hennebelle \& Inutsuka 
(2006) of molecular clouds being 2-phase objects.  
Note that it may be at first surprising to propose a theory for the origin of molecular clouds 
which does not include self-gravity. However, as revealed by CO observations (e.g. 
Bertoldi \& McKee 1992, Heyer et al. 2001),
molecular clouds of mass smaller than $10^3$ solar mass, are not gravitationally bound making very 
hard to invoke gravity as an important physical agent for the formation of these clouds. Heyer et al. 
(2001) also speculate that an external pressure in needed to confine them. External pressure
 is naturally provided by the 2-phase physics.

Finally, we have investigated the distribution of the cloud column density, ${\cal N}(N)$. 
We find that between $N \simeq 2 \times 10^{18}$ cm$^{-2}$ and $N \simeq 2 \times 10^{19}$ cm$^{-2}$,
${\cal N}(N) \propto N^{-1.2}$ whereas from their survey, Heiles \& Troland (2005) inferred 
${\cal N}(N) \propto N^{-1}$ for 
$N \simeq 2 \times 10^{18}$ cm$^{-2}$ and $N \simeq 2 \times 10^{20}$ cm$^{-2}$. 
We speculate that the discrepancy in the range of column density is due to the setup of the
 numerical simulation which limits the highest column densities. We also propose that 
the index of the power law is due to the flow being nearly scale-invariant, i.e. 
${\cal N}(L) \propto L ^{-D}$ ($D$ being the dimension),  and the CNM structures being 
roughly uniform in density. Whereas the first is  likely to be a
consequence of turbulence, the second 
is a consequence of the flow being thermally bistable, that is to say the density 
of the CNM does not change with scale or structure mass.

\begin{acknowledgements}
We acknowledge  the support of  the CEA computing center,  CCRT, where
all the  simulations where  carried out. 
We are very grateful to Snezana Stanimirovi\'c, the referee, for 
an insightful and constructive report which has significantly 
contributed to improve the paper.
It is a pleasure to thank  Carl Heiles 
 and Thomas Troland  for enlightening discussions 
 on various observational aspects of the atomic interstellar medium. 
PH is grateful to Edith Falgarone and Michel P\'erault, for many
discussions on related topics over the years. 
\end{acknowledgements}

\end{document}